\documentclass[11pt]{article}
\usepackage{fullpage}
\usepackage{amsmath,amsfonts,amssymb,amsthm}
\usepackage{enumerate}
\usepackage{stmaryrd}

\usepackage{xcolor}
\colorlet{orangeblue}{orange!90!blue}
\colorlet{blueyellow}{blue!70!yellow}
\colorlet{yellowgreen}{yellow!70!green}

\colorlet{Uniorange}{orange!90!blue}
\definecolor{Unigreen}{RGB}{0,101,83}
\definecolor{Uniyellow}{RGB}{254,206,16}

\usepackage{tikz}
\usetikzlibrary{calc}
\usetikzlibrary{arrows}
\usetikzlibrary{shadows}
\usetikzlibrary{positioning}
\usetikzlibrary{shapes,patterns,snakes}
\usetikzlibrary{decorations.pathmorphing}
\usetikzlibrary{decorations.markings}
\usetikzlibrary{backgrounds}
\usetikzlibrary{trees}
\usepackage{mathrsfs}

\usepackage{tkz-graph}

\usepackage{pgf}
\usepackage{xxcolor}
\usetikzlibrary{arrows,shadows,petri}

\usepackage{fancybox}

\newcommand{\ignore}[1]{}

\newcommand{\etal}{{\em et al.}~}

\newtheoremstyle{postnum}
  {\topsep}
  {\topsep}
  {\slshape}
  {0pt}
  {\bfseries}
  {:}
  { }
  {\thmname{#1}\thmnote{ (#3)}}

\theoremstyle{postnum}

\newtheoremstyle{prenum}
  {\topsep}
  {\topsep}
  {\slshape}
  {0pt}
  {\bfseries}
  {.}
  { }
  {\thmnumber{#2}\thmname{ #1}\thmnote{ (#3)}}

\theoremstyle{prenum}

\newtheorem{theorem}{Theorem}[section]
\newtheorem{definition}[theorem]{Definition}

\newtheorem{corollary}[theorem]{Corollary}
\newtheorem{proposition}[theorem]{Proposition}
\newtheorem{fact}[theorem]{Fact}

\newenvironment{myproblem}[3]
  {\vspace{0.1in} 
	{
	\par\noindent{{\sc #1}}
	\par\noindent{{\em Instance}: #2}
	\par\noindent{{\em Question}: #3}
	}
  {\vspace{0.1in}}}

\newcommand{\ket}[1]{| #1 \rangle}

\newcommand{\ketbra}[2]{| #1 \rangle \langle #2 |}

\renewcommand{\Pr}{\mathbb{P}}
\newcommand{\Exp}{\mathbb{E}}

\newcommand{\norm}[1]{\lVert #1 \rVert}
\newcommand{\ip}[2]{\langle #1 , #2 \rangle}

\newcommand{\ii}{\mathtt{i}}

\newcommand{\Id}{\mathbb{I}}

\newcommand{\RR}{\mathbb{R}}
\newcommand{\CC}{\mathbb{C}}

\newcommand{\Hb}[1]{\mathbb{C}^{#1}}
\newcommand{\Dens}{\mathcal{D}}
\newcommand{\Lin}{\mathrm{L}}
\newcommand{\Sup}{\mathrm{T}}
\newcommand{\Cp}{\mathrm{CP}}
\newcommand{\Pos}{\mathrm{Pos}}
\newcommand{\Ch}{\mathrm{C}}
\newcommand{\Euc}[1]{\mathcal{#1}}

\newcommand{\Acc}{\mathrm{Acc}}
\newcommand{\Val}{\mathrm{V}}

\newcommand{\Baire}{\mathcal{B}}
\newcommand{\Simplex}{\mathcal{P}}

\DeclareMathOperator{\Tr}{Tr}

\DeclareMathOperator{\vspan}{span}

\newif\ifnotesw\noteswtrue
   {\ifnotesw\marginpar[\hfill\(\top\)]{\(\top\)}\fi}%
   {\ifnotesw\marginpar[\hfill\(\bot\)]{\(\bot\)}\fi}

\newcommand{\mnote}[1]%
    {\ifnotesw\marginpar%
        [{\scriptsize\begin{minipage}[t]{\marginparwidth}
        \raggedleft#1%
                        \end{minipage}}]%
        {\scriptsize\begin{minipage}[t]{\marginparwidth}
        \raggedright#1%
                        \end{minipage}}%
    \fi}

\title{A Note on Quantum Markov Models}
\author{Christino Tamon\footnote{Department of Computer Science, Clarkson University, Potsdam, NY, USA 13699. Email: {\tt ctamon@clarkson.edu}.} 
\and
Weichen Xie\footnote{Department of Mathematics, Clarkson University, Potsdam, NY, USA 13699. Email: {\tt xiew@clarkson.edu}.} 
}
\date{\today}
\begin{document}
\maketitle
\begin{abstract}
The study of Markov models is central to control theory and machine learning.
A quantum analogue of partially observable Markov decision process
was studied in (Barry, Barry, and Aaronson, {\em Phys. Rev. A}, 90, 2014).
It was proved that goal-state reachability is undecidable in the quantum setting, whereas it is decidable classically.
In contrast to this classical-to-quantum transition from decidable to undecidable, 
we observe that the problem of approximating the optimal policy which maximizes the average discounted reward 
over an infinite horizon remains decidable in the quantum setting.
Given that most relevant problems related to Markov decision process are undecidable classically
(which immediately implies undecidability in the quantum case), this provides one of the few examples 
where the quantum problem is tractable.
\end{abstract}


\section{Introduction}

The study of Markov models is important to control theory and machine learning. In optimal control, it is relevant 
for the theory of sequential decision making under uncertainty. Seminal results in this area were proved by 
Blackwell \cite{b65}, Sondik \cite{s78}, and others (see \cite{bs78}).
In machine learning, it is central for studying the main problems in reinforcement learning, 
such as planning and optimization (see \cite{rn10,sb18}). 
The primary model used in both areas is the Markov decision process (partially observable or otherwise).

For quantum computing, Markov models are useful for developing algorithms and for modeling the quantum phenomena 
observed in quantum systems. In the design of algorithms, several fundamental algorithms, such as Grover search, 
are based on quantization of Markov chain on graphs (see Szegedy \cite{s04}). 
Further generalizations to quantum Markov processes were studied in quantum information theory 
in the context of quantum channels (see \cite{tkrwv10}). 

In a recent work, Barry \etal \cite{bba14} introduced a model of quantum observable Markov decision process. 
This is a natural quantum analogue of a partially observable Markov decision process, although it is a strict
generalization of a {\em belief} Markov decision process. The latter is a commonly used full-information model 
to represent a partially observable Markov decision process.
The main component used to define a quantum observable Markov decision process is a set of superoperators. 
Using standard machinery from quantum information (see \cite{w18,w17}), we recast the set of superoperators 
as composition of a conditional channel and a quantum instrument. 
We describe this in the context of quantum transducers.

A finite transducer is a finite automata with output (see \cite{hu79}). There are two standard models for transducers:
the Moore machine and the Mealy machine. In the former model, the output is generated from the current state, while
in the second model, the output is generated based on the action taken from the current state.
Following in this classical vein, we describe models for quantum Moore and quantum Mealy machines, and
we prove that they are equivalent. The proof of this equivalence relies on the decoupling of a superoperator
into the composition of a conditional channel and a quantum instrument.
In earlier works, Wiesner \etal \cite{wc08} and Monras \etal \cite{mbw11} had also studied quantum transducers 
but in the context of stochastic generators.

In \cite{bba14}, Barry \etal studied the goal-state perfect reachability problem (also called planning) for
quantum observable Markov decision process. Here, the objective is to find a sequence of actions which 
will evolve the system from the initial state to a given target state, with probability one. 
Their main result is that this perfect planning problem is undecidable in the quantum setting, 
whereas the classical version is known to be decidable.

Madani \etal \cite{mhc03} proved that most classical problems related to 
partially observable Markov decision process, such as planning and optimization, are undecidable.
Since the quantum problems are generalizations of the classical ones, 
this immediately implies that the quantum problems are also undecidable.
The list of these problems do not include the above perfect planning problem.
So, Barry \etal \cite{bba14} shows an interesting classical-to-quantum transition from decidable to undecidable. 
A similar phenomenon occurs for the perfect non-occurrence problem where we ask if there is an output sequence 
that will never be observed. This problem is decidable in the classical setting,
but its quantum version (called the quantum measurement occurrence problem) was proved undecidable 
by Eisert \etal \cite{emg12}.

The policy existence problem in an infinite horizon that are proved undecidable 
by Madani \etal \cite{mhc03} include the problems 
under the total reward, average reward, and discounted reward criteria.
For approximating the optimal policy, the problems under the total and average reward criteria
remain undecidable, but the problem under the discounted criteria is decidable.
We focus on this latter problem and observe that the quantum analogue is still decidable.
This is obtained by revisiting the work of Blackwell \cite{b65} on Markov decision process 
over Borel spaces. We also adapt some observations of Sondik \cite{s78} on the
compact representation of the optimal policy to the quantum setting.

\begin{figure}[t]
\begin{center}
\begin{tabular}{|l||l|l|} \hline 
Problem						&	Classical 						&	Quantum 	\\ \hline \hline
Reachability				&	no	(Madani \etal \cite{mhc03})	&	{\tt no}	\\ 
Perfect Reachability		&	yes (folklore)					&	no	(Barry \etal \cite{bba14}) \\ 
Non-Occurrence				&	{\em no} (Proposition \ref{prop:nobs-undecidable})								
																&	{\tt no} \\ 
Perfect Non-Occurrence		&	yes (Eisert \etal \cite{emg12})	&	no  (Eisert \etal \cite{emg12}) \\ 
Policy-Existence			&	no	(Madani \etal \cite{mhc03})	&	{\tt no}	\\ \hline
Approximate Policy-Existence	&	yes (Blackwell, Sondik \cite{b65,s78})	
																&	{\em yes} (Theorem \ref{thm:qopt-decidable}) \\ \hline 	
\end{tabular}
\caption{(Un)Decidabilities of some problems related to partially or quantum observable Markov decision process. 
The yes/no entries indicate whether the problem is decidable or not.
}
\label{fig:summary}
\end{center}
\end{figure}

We summarize the complexities of the problems related to Markov decision process in Figure \ref{fig:summary}.


\section{Preliminaries}

\subsection{Markov models}

Suppose $S$, $\Sigma$ and $\Delta$ are finite sets (of states, input actions, and output signals, respectively).
Let $\{P_a : a \in \Sigma\}$ be a collection of Markov chains with a common state space $S$,
where each $P_a$ is a $|S| \times |S|$ column-stochastic matrix. We view each column $j$ of $P_a$ as defining a 
conditional probability distribution $P_a(i|j)$ over $S$, where $i,j \in S$.
Let $Q$ be a $|\Delta| \times |S|$ column-stochastic matrix.
We view each column $j$ of $Q$ as defining a conditional probability distribution $Q(b|j)$ over $\Delta$,
where $b \in \Delta$ and $j \in S$.
Let $p_0 \in \Simplex(S)$ be an initial probability distribution over $S$.
Suppose $R$ is a bounded real-valued reward function over $S$ and $\gamma \in [0,1)$ is a discount factor.
The tuple $M=(S,\Sigma,\Delta,\{P_a\},Q,R,p_0,\gamma)$ is called a {\em partially observable Markov decision process} (POMDP).

There are several well-known Markov models that can be derived from the above.
We obtain a Markov chain if $\Sigma$ is a unary alphabet and $R$ is trivial (say, the constant zero function), 
and we let $\Delta = S$ and $Q = \Id$ for the observed chain, or let $\Delta$ be unary (and $Q$ trivial) for the unobserved chain.
We derive a {\em Markov decision process} (MDP) if $\Delta = S$ and $Q = \Id$, and denote this 
simply as $M=(S,\Sigma,\{P_a\},R,p_0,\gamma)$.
We get a {\em hidden Markov model} (HMM) if $\Sigma$ is a unary alphabet and $R$ is trivial, 
and denote this as $M=(S,\Delta,P,Q,p_0)$.

The random process induced by a partially observable Markov decision process is defined as follows.
For a sequence $(a_n)$, where $a_n \in \Sigma$, we consider the sequence of pairs of random variables 
$(X_n,Y_n)_{n=0}^{\infty}$, $X_n \in S$ and $Y_n \in \Delta$, defined so that:
\begin{equation}
\Pr[X_0 = i] = (p_0)_i,
\ \hspace{.2in} \ 
\mbox{for $i \in S$}
\end{equation}
\begin{equation}
\Pr[X_{n+1}=i |X_n=j] = (P_{a_n})_{ij},
\ \hspace{.2in} \ 
\mbox{for $i,j \in S$}
\end{equation}
\begin{equation}
\Pr[Y_n = k | X_n=j] = Q_{kj},
\ \hspace{.2in} \ 
\mbox{for $k \in \Delta$, $j \in S$}
\end{equation}
A {\em policy} is a sequence $\pi=(\pi_n)_{n=1}^{\infty}$ of maps, 
where $\pi_{n+1}: \Simplex(S) \times H_{n} \rightarrow \Sigma$ 
and $H_n: (\Delta \times \Sigma)^{n} \times \Delta$ is the history at time $n$, for each $n=0,1,\ldots$. 
Thus, if $h_n = (y_0,a_1,\ldots,y_{n-1},a_n,y_n)$ is the history at time $n$, 
then $a_{n+1} = \pi_{n+1}(p_0,h_{n})$ is the action taken, for all $n$.
The value function of $\pi$ is given by
\begin{equation}
\Val_\pi(p_0) = \Exp\left[ \sum_{i=0}^{\infty} \gamma^i R(X_i) \right].
\end{equation}
The expectation is taken over an infinite product probability space\footnote{A brief discussion on the existence 
of this infinite probability space is given in Appendix \ref{section:inf-product}}.
%
The main objective is to compute the optimal value function
$V^\star(p) = \sup_\pi \Val_\pi(p)$,
and to find $\pi^\star$ for which the maximum is achieved.
A policy $\pi = (\pi_n)$ is called {\em stationary} if there is a map 
$f: \Simplex(S) \times \Delta \rightarrow \Sigma$ so that $\pi_n = f$ for all $n$.

\subsection{Quantum information}

We briefly review some basic notation and background from quantum information 
(see Nielsen and Chuang \cite{nc00}, Watrous \cite{w18}).
For a finite set $X$, let $\Euc{X} = \Hb{X}$ denote the complex Euclidean space spanned by the unit vectors 
$\{\ket{a} : a \in X\}$. We will use the standard Dirac notation throughout. 
It will often be convenient to view $\Euc{X}$ as a Hilbert space.

Given complex Euclidean spaces $\Euc{X}$ and $\Euc{Y}$, let $\Lin(\Euc{X},\Euc{Y})$ be the set of linear operators 
from $\Euc{X}$ to $\Euc{Y}$. 
For an operator $A \in \Lin(\Euc{X},\Euc{Y})$, let $A^\dagger$ denote the unique operator in $\Lin(\Euc{Y},\Euc{X})$ 
for which $\ip{v}{Au} = \ip{A^\dagger v}{u}$ for all $u \in \Euc{X}$ and $v \in \Euc{Y}$.
For brevity, we use $\Lin(\Euc{X})$ when $\Euc{X} = \Euc{Y}$.
We say an operator $A \in \Lin(\Euc{X})$ is positive semidefinite if $A = B^\dagger B$ for some $B \in \Lin(\Euc{X})$.
The set of all positive semidefinite operators in $\Lin(\Euc{X})$ is denoted $\Pos(\Euc{X})$.
A positive semidefinite operator $A$ is called a density matrix if $\Tr(A)=1$. 
The set of all density matrices over $\Euc{X}$ is denoted $\Dens(\Euc{X})$.
These density matrices will provide a convenient representation for quantum states.

Let $\Sup(\Euc{X},\Euc{Y})$ be the set of linear operators from $\Lin(\Euc{X})$ to $\Lin(\Euc{Y})$. 
An operator $S \in \Sup(\Euc{X},\Euc{Y})$ is called {\em positive} if it maps a positive semidefinite $A \in \Pos(\Euc{X})$
to a positive semidefinite $S(A) \in \Pos(\Euc{Y})$. 
An operator $S \in \Sup(\Euc{X},\Euc{Y})$ is called {\em completely positive} if $S \otimes \Id_{\Euc{Z}}$ 
is positive for every auxiliary space $\Euc{Z}$.
The set of all linear operators in $\Sup(\Euc{X},\Euc{Y})$ which are completely positive is denoted $\Cp(\Euc{X},\Euc{Y})$.
An operator $S \in \Sup(\Euc{X},\Euc{Y})$ is called {\em trace preserving} if $\Tr(S(A)) = \Tr(A)$ for all $A \in \Lin(\Euc{X})$.
The set of linear operators in $\Sup(\Euc{X},\Euc{Y})$ that are completely positive and trace preserving is denoted $\Ch(\Euc{X},\Euc{Y})$.
Each element of $\Ch(\Euc{X},\Euc{Y})$ is called a quantum {\em channel}.
These quantum channels will provide a representation of quantum operations.

For a finite alphabet $\Sigma$, a {\em conditional channel} over $\Sigma$ is a collection of channels
$\{\Phi_a \in \Ch(\Euc{X}) : a \in \Sigma\}$.
The input to a conditional channel is a classical-quantum state $\ketbra{a}{a} \otimes \rho$, where
$a \in \Sigma$ and $\rho \in \Dens(\Euc{X})$.
First, the conditional channel prepares the following quantum state:
\begin{equation}
\Phi(\rho) = \frac{1}{|\Sigma|}\sum_{b \in \Sigma} \ketbra{b}{b} \otimes \Phi_b(\rho).
\end{equation}
It then applies the projection $\ketbra{a}{a} \otimes \Id$ and returns the second register.

For a finite alphabet $\Delta$, a {\em quantum instrument} $\Omega$ over $\Delta$
is a collection $\{\Omega_b \in \Cp(\Euc{X}) : b \in \Delta\}$ of completely positive operators
for which $\sum_{b \in \Delta} \Omega_b$ forms a channel.
On input $\rho \in \Dens(\Euc{X})$, the instrument first prepares the following quantum state:
\begin{equation}
	\Omega(\rho) = \sum_{b \in \Delta} \ketbra{b}{b} \otimes \Omega_b(\rho).
\end{equation}
Note that $\Omega$ is a quantum channel.
Then, the quantum instrument applies the measurement $\mu: \Delta \rightarrow \Pos(\Euc{X})$ where
$\mu(b) = \ketbra{b}{b} \otimes \Id$, for each $b \in \Delta$.
Here, an element $b \in \Delta$ is selected at random with probability
$p(b) = \ip{\mu(b)}{\Omega(\rho)} = \Tr(\Omega_b(\rho))$ and, 
conditioned on the measurement outcome $b$ was observed, 
the post-measurement state is $\Omega_b(\rho)/\Tr(\Omega_b(\rho))$
(otherwise, if the outcome was not observed, the resulting state is $\sum_b \Omega_b(\rho)$).


\section{Quantum Transducers}

A transducer is a finite automata with output (see Hopcroft and Ullman \cite{hu79}).
We consider two standard models of transducers, namely Moore and Mealy machines, 
and propose their natural quantum analogues. Then, we show that the two quantum models are equivalent.

In this and subsequent sections, we assume $S$ is a finite set of states, $\Sigma$ is a finite set of input symbols, 
and $\Delta$ is a finite set of output symbols. 
We also assume that $\Euc{S} = \Hb{S}$ is a complex Euclidean space associated with $S$.

A {\em quantum Moore machine} is a tuple $M = (S,\Sigma,\Delta,\Phi,\Omega,\rho_0,\Pi)$ where
$\Phi = \{\Phi_a \in \Ch(\Euc{S}) : a \in \Sigma\}$ is a {conditional channel}, 
$\Omega = \{\Omega_b \in \Cp(\Euc{S}) : b \in \Delta\}$ is a {quantum instrument},
$\rho_0$ is the initial density matrix,
and $\Pi$ is an orthogonal projection (onto an accepting subspace).
A single step of $M$ is a composition of the transition map $\Phi$ followed by the output map $\Omega$.
If the current state of $M$ is given by the density matrix $\rho_0$ and the current input is $a \in \Sigma$,
then $M$ transforms $\rho_0$ to the intermediate state $\rho_1 = \Phi_a(\rho)$. Further, $M$ applies the instrument 
$\Omega$ to $\rho_1$ and generates an output $b \in \Delta$ with probability $\Tr\Omega_b(\rho_1)$. The
final resulting state is $\rho_2 = \Omega(\rho_1)/\Tr(\Omega(\rho_1))$.

For input $\alpha = a_1\ldots a_n \in \Sigma^\star$ and output $\beta = b_1\ldots b_n \in \Delta^\star$,
the probability that $M$ on input $\alpha$ will output $\beta$ is denoted $p_M(\beta;\alpha)$,
while the final resulting state of $M$ is denoted $\rho_M(\alpha,\beta)$.
The function computed by $M$ on input $\alpha$ and output $\beta$ is given by
\begin{equation}
\Acc_M(\alpha,\beta) = \Tr(\Pi \rho_M(\alpha,\beta)).
\end{equation}
This is the probability that the final state belongs to the accepting subspace.
The function computed by $M$ on input $\alpha$ is given by
\begin{equation}
\Acc_M(\alpha) = \Tr(\Pi \rho_M(\alpha))
\end{equation}
where
\begin{equation}
\rho_M(\alpha) = \sum_{\beta} p_{M}(\beta;\alpha) \rho_M(\alpha,\beta).
\end{equation}

\begin{fact} \label{fact:moore-properties}
Let $M = (S,\Sigma,\Delta,\Phi,\Omega,\rho_0,\Pi)$ be a quantum Moore machine.
For input $\alpha = a_1\ldots a_n \in \Sigma^\star$ and output $\beta = b_1\ldots b_n \in \Delta^\star$, we have
\begin{equation}
\rho_M(\alpha,\beta) = \frac{\Omega_{b_n} \circ \Phi_{a_n} \circ \ldots \circ \Omega_{b_1} \circ \Phi_{a_1}(\rho_0)}{p_M(\beta;\alpha)},
\ \ \
p_M(\beta;\alpha) = \Tr(\Omega_{b_n} \circ \Phi_{a_n} \circ \ldots \circ \Omega_{b_{1}} \circ \Phi_{a_{1}}(\rho_0)).
\end{equation}

\begin{proof} 
We prove this by induction on the length $n = |\alpha| = |\beta|$.
For $n=1$, the state after the conditional channel is $\tilde{\rho}_1 = \Phi_{a_1}(\rho_0)$.
By the properties of the quantum instrument, the probability that $M$ outputs $b_1$ is given by
\begin{equation}
p_M(b_1 ; a_1) = \Tr(\Omega_{b_1}(\tilde{\rho}_1)) = \Tr(\Omega_{b_1} \circ \Phi_{a_1}(\rho_0))
\end{equation}
and the post-measured state of
\begin{equation}
\rho_M(a_1,b_1) = \frac{\Omega_{b_1} \circ \Phi_{a_1}(\rho_0)}{p_M(b_1 ; a_1)}.
\end{equation}
This proves the base case.

Assume the claim holds for $n \ge 1$. 
Suppose the input is $\alpha = a_1\ldots a_{n+1}$ and the output is $\beta = b_1\ldots b_{n+1}$.
Let $\rho_n = \rho_M(a_1\ldots a_n, b_1\ldots b_n)$ and $\tilde{\rho}_{n+1} = \Phi_{a_{n+1}}(\rho_{n})$.
Then, the probability that $M$ outputs $b_{n+1}$ on input $a_{n+1}$ given that it has read input $a_1\ldots a_n$
and emitted output $b_1\ldots b_n$ is
\begin{equation}
\Tr(\Omega_{b_{n+1}}(\tilde{\rho}_{n+1}))
	= \Tr(\Omega_{b_{n+1}} \circ \Phi_{a_{n+1}}(\rho_{n})) \\
	= \frac{\Tr(\Omega_{b_{n+1}} \circ \Phi_{a_{n+1}} \circ \ldots \circ \Omega_{b_1} \circ \Phi_{a_1}(\rho_{0}))}{p_M(b_1 \ldots b_n ; a_1 \ldots a_n)}
\end{equation}
where the last step follows from the inductive hypothesis.
Therefore,
\begin{equation}
p_M(b_1 \ldots b_{n+1} ; a_1 \ldots a_{n+1}) 
	= \Tr(\Omega_{b_{n+1}} \circ \Phi_{a_{n+1}} \circ \ldots \circ \Omega_{b_{1}} \circ \Phi_{a_{1}}(\rho_0)).
\end{equation}
The resulting quantum state is
\begin{equation}
\rho_{n+1} 
	= \frac{\Omega_{b_{n+1}} \circ \Phi_{a_{n+1}}(\rho_{n})}{\Tr(\Omega_{b_{n+1}} \circ \Phi_{a_{n+1}}(\rho_{n}))}
	= \frac{\Omega_{b_{n+1}} \circ \Phi_{a_{n+1}} \circ \ldots \circ \Omega_{b_1} \circ \Phi_{a_1}(\rho_0)}
			{\Tr(\Omega_{b_{n+1}} \circ \Phi_{a_{n+1}} \circ \ldots \circ \Omega_{b_1} \circ \Phi_{a_1}(\rho_0))}.
\end{equation}
This shows the claim holds for $n+1$.
\end{proof}
\end{fact}

A {\em quantum Mealy machine} is a tuple $M = (S,\Sigma,\Delta,\Lambda,\rho_0,\Pi)$ where
$\Lambda = \{\Lambda_a : a \in \Sigma\}$ is a set of quantum instruments over a common alphabet $\Delta$,
$\rho_0$ is the initial density matrix, and
$\Pi$ is an orthogonal projection (onto an accepting subspace).
Each quantum instrument $\Lambda_a$ is a composition of a channel and a complete measurement over $\Delta$.
For notational convenience, we will identify this channel $\Lambda_a$ as well (it becomes an instrument after 
the first register is measured; see Watrous \cite{w18}).
More specifically, we associate $\Lambda_a$ with a collection 
$\{\Lambda_{a,b} : b \in \Delta\} \subseteq \Cp(\Euc{S})$ of completely positive operators
satisfying $\sum_{b \in \Delta} \Lambda_{a,b} \in \Ch(\Euc{S})$.
Then, we define the channel
\begin{equation}
	\Lambda_a(\rho) = \sum_{b \in \Delta} \ketbra{b}{b} \otimes \Lambda_{a,b}(\rho).
\end{equation}
Upon measuring the first register, the output is $b$ with probability $\Tr(\Lambda_{a,b}(\rho))$
and the post-measurement state is $\Lambda_{a,b}(\rho)/\Tr(\Lambda_{a,b}(\rho))$.
Thus, if the current state of $M$ is $\rho$ and the current input is $a \in \Sigma$, $M$ applies
$\Lambda_a$ to $\rho$ and generates output $b \in \Delta$ according to the above process.

For input $\alpha = a_1\ldots a_n \in \Sigma^\star$ and output $\beta = b_1\ldots b_n \in \Delta^\star$,
the probability that $M$ on input $\alpha$ will output $\beta$ is denoted $p_M(\beta;\alpha)$,
while the final resulting state of $M$ is denoted $\rho_M(\alpha,\beta)$.
The function computed by $M$ on input $\alpha$ with output $\beta$ is given by
\begin{equation}
\Acc_M(\alpha,\beta) = \Tr(\Pi \rho_M(\alpha,\beta)).
\end{equation}
This is the probability that the final state belongs to the accepting subspace.
The function computed by $M$ on input $\alpha$ is given by
\begin{equation}
\Acc_M(\alpha) = \Tr(\Pi \rho_M(\alpha))
\end{equation}
where
\begin{equation}
\rho_M(\alpha) = \sum_{\beta} p_{M}(\beta;\alpha) \rho_M(\alpha,\beta).
\end{equation}

\begin{fact} \label{fact:mealy-properties}
Let $M = (S,\Sigma,\Delta,\Lambda,\rho_0,\Pi)$ be a quantum Mealy machine,
where $\Lambda = \{\Lambda_{a,b} \in \Cp(\Euc{S}) : a \in \Sigma, b \in \Delta\}$
so that for each $a \in \Sigma$, the collection $\{\Lambda_{a,b} : b \in \Delta\}$ forms an instrument.
Then, for any input $\alpha = a_1\ldots a_n \in \Sigma^\star$ and 
output $\beta = b_1\ldots b_n \in \Delta^\star$, we have
\begin{equation}
\rho_M(\alpha,\beta) =
	\frac{\Lambda_{a_n,b_n} \circ \ldots \circ \Lambda_{a_1,b_1}(\rho_0)}{p(\beta ; \alpha)},
\ \ \
p(\beta ; \alpha) = \Tr(\Lambda_{a_n,b_n} \circ \ldots \circ \Lambda_{a_1,b_1}(\rho_0)).
\end{equation}
\end{fact}


\begin{definition}
Let $M_1$ be a quantum Moore machine and $M_2$ be a quantum Mealy machine sharing the
same input alphabet $\Sigma$ and output alphabet $\Delta$.
We say $M_1$ and $M_2$ are {\em equivalent} if for all inputs $\alpha \in \Sigma^\star$ and
outputs $\beta \in \Delta^\star$, we have 
$p_{M_1}(\beta ; \alpha) = p_{M_2}(\beta ; \alpha)$
and
$\Acc_{M_1}(\alpha,\beta) = \Acc_{M_2}(\alpha,\beta)$.
\end{definition}

In what follows, we show that the two models of quantum transducers are equivalent.
The idea behind this is the decoupling of the set of quantum instruments in a quantum Mealy machine
into a composition of a conditional channel and a shared quantum instrument in a quantum Moore machine
(see Figure \ref{fig:decouple}).

\begin{proposition} \label{prop:moore2mealy}
For any quantum Moore machine, there is an equivalent quantum Mealy machine. 

\begin{proof}
Let $M_1 = (S,\Sigma,\Delta,\Phi,\Omega,\rho_0,\Pi)$ be a quantum Moore machine.
Consider a quantum Mealy machine $M_2 = (S,\Sigma,\Delta,\Lambda,\rho_0,\Pi)$ where
the collection of instruments $\Lambda = \{\Lambda_{a,b} : a \in \Sigma, b \in \Delta\}$
are defined as
\begin{equation}
\Lambda_{a,b} = \Omega_b \circ \Phi_a.
\end{equation}
Since both $\Phi_a$ and $\Omega_b$ are completely positive, so is $\Lambda_{a,b}$.
Note that for each $a \in \Sigma$, we have
\begin{equation}
\sum_{b \in \Delta} \Lambda_{a,b} = (\sum_{b \in \Delta} \Omega_b) \circ \Phi_a
\end{equation}
is a channel since channels are closed under composition.
Therefore, the following map $\Lambda_a$ is also a channel:
\begin{equation}
\Lambda_a(\rho) 
	= \sum_{b \in \Delta} \ketbra{b}{b} \otimes \Lambda_{a,b}(\rho)
	= \sum_{b \in \Delta} \ketbra{b}{b} \otimes \Omega_b \circ \Phi_a(\rho).
\end{equation}
This is the same state that is prepared by $M_1$ before the measurement.

By Facts \ref{fact:moore-properties} and \ref{fact:mealy-properties}, 
for all $\alpha \in \Sigma^\star$ and $\beta \in \Delta^\star$, we have
$p_{M_1}(\beta ; \alpha) = p_{M_2}(\beta ; \alpha)$
and
$\Acc_{M_1}(\alpha,\beta) = \Acc_{M_2}(\alpha,\beta)$ (since the states agree).
\end{proof}
\end{proposition}


\tikzstyle{boxu}=[draw, minimum size=2em]
\tikzstyle{boxc}=[draw, minimum size=2em]
\tikzstyle{boxq}=[draw, minimum size=2em]
\tikzstyle{box2}=[draw, minimum size=2em]
\tikzstyle{init} = [pin edge={to-,thin,double,black}]
\tikzstyle{exit} = [pin edge={-to,thin,double,black}]

\begin{figure}[t]
\begin{center}
\begin{tikzpicture}[scale=2.0,node distance=1.5cm,auto,>=latex']
    \node [boxu, pin={[exit]below:{\scriptsize signal $b$}}] (a) {{\color{black} $\Lambda_a$}};
    \node (b) [left of=a,node distance=1.3cm, coordinate] {a};
    \node [coordinate] (end) [right of=a, node distance=1.3cm]{};
    \path[->,thick] (b) edge node {{\scriptsize $\rho_t$}} (a);
    \draw[->,thick] (a) edge node {{\scriptsize $\rho_{t+1}$}} (end) ;
\end{tikzpicture}
\quad \quad \quad \quad
\begin{tikzpicture}[node distance=1.5cm,auto,>=latex']
    \node [boxc, pin={[init]below:{\scriptsize action $a$}}] (a) {{\color{black} $\Phi$}};
    \node (b) [left of=a,node distance=1.2cm, coordinate] {a};
    \node [boxq, pin={[exit]below:{\scriptsize signal $b$}}] (x) [right of=a] {{\color{black} $\Omega$}};
    \node [coordinate] (end) [right of=x, node distance=1.2cm]{};
    \path[->,thick] (b) edge node {{\scriptsize $\rho_t$}} (a);
    \path[->,thick] (a) edge node {} (x);
    \draw[->,thick] (x) edge node {{\scriptsize $\rho_{t+1}$}} (end) ;
\end{tikzpicture}
\caption{Decoupling a superoperator into composition of a conditional channel and a quantum instrument.}
\label{fig:decouple}
\end{center}
\end{figure}
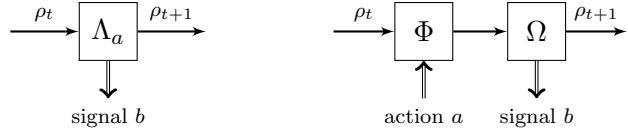

\begin{proposition} \label{prop:mealy2moore}
For any quantum Mealy machine, there is an equivalent quantum Moore machine. 

\begin{proof}
Let $M_1 = (S,\Sigma,\Delta,\Lambda,\rho_0,\Pi)$ be a quantum Mealy machine where
$\Lambda = \{\Lambda_{a,b} \in \Cp(\Euc{S}) : a \in \Sigma, b \in \Delta\}$ so that
$\sum_{b \in \Delta} \Lambda_{a,b} \in \Ch(\Euc{S})$, for each $a \in \Sigma$.

Consider a quantum Moore machine $M_2 = (S,\Sigma,\Delta,\Phi,\Omega,\rho_0,\Pi)$ defined as follows.
We let $\Phi_a$ be a channel given by
\begin{equation}
\Phi_a(\rho) = \sum_{b \in \Delta} \ketbra{b}{b} \otimes \Lambda_{a,b}(\rho)
\end{equation}
and let $\Omega_b$ be a linear operator given by
\begin{equation}
\Omega_b = \ketbra{b}{b} \otimes \Id.
\end{equation}
Note that $\Omega_b$ is completely positive and $\sum_{b \in \Delta} \Omega_b$ is the identity channel. 
Hence, $\{\Omega_b : b \in \Delta\}$ forms a quantum instrument.

We see that
\begin{equation}
(\Omega_b \circ \Phi_a)(\rho) = \ketbra{b}{b} \otimes \Lambda_{a,b}(\rho)
\end{equation}
and
\begin{equation}
\sum_{b \in \Delta} (\Omega_b \circ \Phi_a)(\rho) = \sum_{b \in \Delta} \ketbra{b}{b} \otimes \Lambda_{a,b}(\rho)
\end{equation}
This is the same state that is prepared by $M_1$ before the measurement.

By Facts \ref{fact:moore-properties} and \ref{fact:mealy-properties}, 
for all $\alpha \in \Sigma^\star$ and $\beta \in \Delta^\star$, we have
$p_{M_1}(\beta ; \alpha) = p_{M_2}(\beta ; \alpha)$
and
$\Acc_{M_1}(\alpha,\beta) = \Acc_{M_2}(\alpha,\beta)$ (since the states agree).
\end{proof}
\end{proposition}

By combining Propositions \ref{prop:moore2mealy} and \ref{prop:mealy2moore}, we have the following.

\begin{theorem} \label{thm:moore-is-mealy}
Quantum Mealy and Moore machines are equivalent.
\end{theorem}

\bigskip
\par\noindent{\em Remark}:
The classical transducers are special cases of the quantum transducers.
For example, a Moore machine can be defined as a tuple $M=(S,\Sigma,\Delta,P,Q,s_0,F)$
where $P=\{P_a : a \in \Sigma\}$ is a set of Markov chains, $Q=\{Q_s : s \in S\}$ is a set of probability distributions 
over $\Delta$, $s_0$ is the initial state and $F \subset S$ is a collection of accepting states. 
A Mealy machine is defined similarly except for one difference: the set $Q$ is given by $\{Q_{s,a} : s \in S,a \in \Sigma\}$, 
that is, each probability distribution depends on both the current state and the action taken.
These differ slightly from the models defined in \cite{hu79} in that we retain the notion of accepting states
(which is standard for finite automata).

\bigskip
\par\noindent{\em Remark}:
Note that any Markov chain, hidden Markov model, or partially observable Markov decision process
(and hence Markov decision process) can be simulated by a quantum Moore machine.
Moreover, the decoupled structure of a quantum Moore machine allows us to either focus on 
the input side (conditional channel) or the output side (quantum instrument).
Thus, it is easy to see that a quantum Moore machine can simulate any quantum Markov chain, 
any quantum automata \cite{bjkp05}, and any hidden quantum Markov model \cite{wc08,mbw11}.


\section{Quantum observable Markov decision processes}

We consider the quantum observable Markov decision process defined by Barry \etal \cite{bba14}. 
Let $S$ be a finite set of states, $\Sigma$ be a finite set of input symbols, 
and $\Delta$ be a finite set of output symbols. 
Let $\Euc{S} = \Hb{S}$ be the complex Euclidean space associated with the set of states.

\begin{definition} \label{def:superoperator}
(Eisert \etal \cite{emg12}, Barry \etal \cite{bba14}) \\
A {\em superoperator} $\Lambda$ is given by a set of Kraus operators $\{K_b \in \Lin(\Euc{S})\}_{b=1}^{m}$ 
where $\sum_{b=1}^{m} K_b^\dagger K_b = \Id$. 
If $\Lambda$ is applied to a density matrix $\rho$, then with probability $p_b = \Tr(K_b \rho K_b^\dagger)$, 
the resulting state is $K_b \rho K_b^\dagger/p_b$ and the observation $b$ is emitted.
\end{definition}

We note that a superoperator {\em is} a quantum instrument. 

\begin{definition} \label{def:qomdp}
(Barry \etal \cite{bba14}) \\
A {\em quantum observable Markov decision process} (QOMDP) is a tuple $M=(\Euc{S},\Sigma,\Delta,\Lambda,R,\gamma,\rho_0)$ where
$\Lambda = \{\Lambda_a : a \in \Sigma\}$ is a set of superoperators, 
$R = \{R_a : a \in \Sigma\}$ is a set of operators,
$\rho_0 \in \Euc{S}$ is the initial state,
and $\gamma \in [0,1)$ is a discount factor.
Moreover, we assume each superoperator $\Lambda_a$ is defined by the set of Kraus operators 
$\{\Lambda_{a,b} : b \in \Delta\}$ satisfying $\sum_{b \in \Delta} \Lambda_{a,b}^\dagger\Lambda_{a,b} = \Id$.
The reward associated taking action $a \in \Sigma$ from state $\rho \in \Dens(\Euc{S})$ is given by 
$R(\rho,a) = \Tr(R_a \rho)$.
\end{definition}

The set of superoperators in Definition \ref{def:qomdp} functionally acts as a ``conditional quantum instrument.''
Using Theorem \ref{thm:moore-is-mealy}, we formalize this using conditional channel and quantum instrument
(as defined in Watrous \cite{w18} and in Wilde \cite{w17}).

\medskip
\par\noindent{\em Remark}: The reward operator $R$ must be Hermitian 
since we require the rewards be real for the purpose of optimization.

\begin{fact} \label{fact:bdd-reward}
The reward $R$ is a bounded real-valued function.

\begin{proof}
Suppose $\Euc{S} = \CC^S$ where $S$ is a set of $d$ states. 
So, $R_a$ is a $d \times d$ Hermitian matrix, for all $a$, and $\Tr(R_a\rho) \le d\lambda_{max}(R_a)$, 
where $\lambda_{max}(R_a)$ is the maximum eigenvalue (in absolute value) of $R_a$. 
\end{proof}
\end{fact}

The next proposition shows we may assume without loss of generality that the reward function depends on the
current state alone. We adopt this simplifying assumption from here on.

\begin{proposition} \label{prop:state-based-reward}
Let $M_1=(\Euc{S}_1,\Sigma,\Delta,\Lambda,\{R_a\},\rho_0,\gamma)$ be a quantum observable Markov decision process 
where the reward is defined as $R_1(\rho,a) = \Tr(R_a\rho)$ for all $\rho \in \Dens(\Euc{S}_1)$.
There is an equivalent quantum observable Markov decision process $M_2=(\Euc{S}_2,\Sigma,\Delta,\Gamma,R,\mu_0,\gamma)$
where the reward is defined as $R_2(\mu) = \Tr(R\mu)$ for all $\mu \in \Dens(\Euc{S}_2)$.

\begin{proof}
Suppose $S_1$ is the underlying finite set of states of $M_1$. Let $S_2 = S_1 \times \Sigma$. 
Since $\CC^{S_2} = \CC^{S_1} \otimes \CC^\Sigma$, we have $\Euc{S}_2 = \Euc{S}_1 \otimes \CC^\Sigma$.
Let $\mu_0 = \rho_0 \otimes \ketbra{a_0}{a_0}$, for some arbitrary $a_0 \in \Sigma$.

Each quantum instrument $\Gamma_a$ of $\Gamma$ is assumed to act on product states and is defined as
\begin{equation}
\Gamma_a(\rho \otimes \alpha) = \Lambda_a(\rho) \otimes \ketbra{a}{a}.
\end{equation}
So, $\Gamma_a$ applies the quantum instrument $\Lambda_a$ of $\Lambda$ to its first component 
and applies the replacement channel $\alpha \mapsto \Tr(\alpha)\ketbra{a}{a}$ to its second component.

We define $R = \sum_{b \in \Sigma} R_b \otimes \ketbra{b}{b}$. Then,
$R_2(\rho \otimes \ketbra{a}{a}) 
	= \Tr(R_a\rho \otimes \ketbra{a}{a})
	= R_1(\rho,a)$.
So, if $(\rho_n)_{n=0}^{\infty}$ is the state trajectory of $M_1$ with a sequence of actions $(a_n)_{n=0}^{\infty}$,
then the corresponding trajectory of $M_2$ is $(\rho_n \otimes \ketbra{a_n}{a_n})_{n=0}^{\infty}$.
Thus,
$\Exp\left[\sum_{i} \gamma^i R_1(\rho_i,a_i)\right] = \Exp\left[\sum_{i} \gamma^i R_2(\rho_i \otimes \ketbra{a_i}{a_i})\right]$.
\end{proof}
\end{proposition}

Suppose $M = (\Euc{S},\Sigma,\Delta,\Lambda,R,\gamma,\rho_0)$ is a quantum observable Markov decision process.
Given a sequence of actions $\alpha = a_1\ldots a_n \in \Sigma^\star$ and a sequence of observations 
$\beta = b_1\ldots b_n \in \Delta^\star$, we have
\begin{equation}
\rho_M(\alpha, \beta)
	= \frac{\Lambda_{a_n,b_n}\ldots\Lambda_{a_1,b_1}\rho_0\Lambda_{a_1,b_1}^\dagger\ldots\Lambda_{a_n,b_n}^\dagger}{p_M(\beta;\alpha)},
\ \ \
p_M(\beta;\alpha) 
	= {\Tr(\Lambda_{a_n,b_n}\ldots\Lambda_{a_1,b_1}\rho_0\Lambda_{a_1,b_1}^\dagger\ldots\Lambda_{a_n,b_n}^\dagger)}
\end{equation}
where $\rho_M(\alpha,\beta)$ is the state of $M$ on input $\alpha$ and output $\beta$ 
and $p_M(\alpha,\beta)$ is the probability that $M$ outputs $\beta$ on input $\alpha$.

\begin{definition} \label{def:goal-qomdp}
(Barry \etal \cite{bba14}) \\
A {\em goal quantum observable Markov decision process} is a tuple $M=(\Euc{S},\Sigma,\Delta,\Lambda,\rho_0,\rho_g)$ 
where $\Euc{S}$, $\Sigma$, $\Delta$, $\Lambda$, and $\rho_0$ are as defined for quantum observable Markov decision process,
and $\rho_g$ is the goal state. The goal state is absorbing, that is, for all $a \in \Sigma$ and $b \in \Delta$, 
if $p_{a,b} = \Tr(\Lambda_{a,b}\rho_g\Lambda_{a,b}^\dagger) > 0$,
then $\Lambda_{a,b}\rho_g\Lambda_{a,b}^\dagger/p_{a,b} = \rho_g$.
\end{definition}

\begin{fact}
For every goal quantum observable Markov decision process, there are equivalent quantum Mealy 
and quantum Moore machines.

\begin{proof}
(Sketch) We let the Mealy projector be a projection onto the one-dimensional subspace spanned by the unique goal state
of the goal quantum observable Markov decision process.
The equivalent quantum Moore machine is obtained from the quantum Mealy machine using Theorem \ref{thm:moore-is-mealy}.
\end{proof}
\end{fact}

Similar to Definition \ref{def:goal-qomdp}, we may also define a {\em goal partially observable Markov decision process} 
as $M=(S,\Sigma,\Delta,\{P_a\},Q,s_0,s_g)$ where $s_g$ is an absorbing state for each Markov chain $P_a$, $a \in \Sigma$.
Likewise, we can show that $M$ is equivalent to a Moore machine.
Note that the structure of a goal partially observable Markov decision process 
(used in, for example, Madani \etal \cite{mhc03}) is closer to the structure of a Moore machine 
than to a Mealy machine (see Hopcroft and Ullman \cite{hu79}, page 42-43).
This is due to the decoupled nature of the set $P$ of Markov chains and the stochastic output function $Q$ 
in a goal partially observable Markov decision process.


\section{Complexity}

We explore some computational problems related to partially observable and quantum observable Markov decision processes. 
These include reachability (or planning) and occurrence problems.

\begin{myproblem}
{{Reachability}}
{A Moore machine $M = (S,\Sigma,\Delta,\{P_a\},Q,s_0,s_g)$, where $s_g \in S$ is an absorbing goal state, 
and a threshold $\tau \in [0,1]$.}
{Is there $\alpha \in \Sigma^\star$ so that $\Acc_M(\alpha) \ge \tau$?}
\end{myproblem}

The above problem is called the {\em plan-existence} or {\em probabilistic planning} problem in \cite{mhc03}.

\begin{theorem} \label{thm:classical-planning} (Madani, Hanks, and Condon \cite{mhc03}, Theorem 3.7) \\
{\sc Reachability} is undecidable.
\end{theorem}

\par\noindent
In contrast, the reachability problem with $\tau = 1$ is decidable (a proof was given in \cite{bba14}). 

\begin{myproblem}
{{Quantum Reachability}}
{A quantum Moore machine $M = (\Euc{S},\Sigma,\Delta,\Phi,\Omega,\rho_0,\Pi)$ and a threshold $\tau \in [0,1]$.}
{Is there $\alpha \in \Sigma^\star$ so that $\Acc_M(\alpha) \ge \tau$?}
\end{myproblem}

\begin{theorem} \label{thm:qreach} 
{\sc Quantum Reachability} is undecidable.

\begin{proof}
This follows from Theorem \ref{thm:classical-planning} since a quantum Moore machine can simulate a Moore machine.
\end{proof}
\end{theorem}

An alternative proof of Theorem \ref{thm:qreach} was also given by Wolf \etal \cite{wcp11}.

\begin{theorem} (Barry, Barry, and Aaronson \cite{bba14}) \\
{\sc Quantum Reachability} with $\tau=1$ is undecidable.
\end{theorem}

Blondel \etal \cite{bjkp05} proved that the quantum reachability problem where the conditional channel 
consists of unitary channels with a trivial quantum instrument 
is decidable if and only if the threshold condition is a strict inequality.

\bigskip
In the occurrence problem, the goal is to determine if, in a partially observable Markov decision process, 
the probability of observing a sequence of output is less or equal to a given threshold. 
Informally, this asks if the output sequence is a rare anomaly.

\begin{myproblem}
{{Non-Occurrence}}
{A Moore machine $M = (S,\Sigma,\Delta,\{P_a\},Q,p_0)$ and $\tau \in [0,1]$.}
{Are there are $\alpha \in \Sigma^\star$ and $\beta \in \Delta^\star$ so that $\Acc_M(\alpha,\beta) \le \tau$?}
\end{myproblem}

Although the following result is possibly folklore, for completeness, we provide a simple proof.

\begin{proposition} \label{prop:nobs-undecidable}
{\sc Non-Occurrence} is undecidable.

\begin{proof}
We reduce goal-state reachability to non-occurrence.
Let $M_1=(S_1,\Sigma,\Delta_1,\{P_a\},Q_1,p_0,s_g)$ be a Moore machine and $\tau \in (0,1)$ is a threshold. 
Recall that $P_a(j,i)$ is the probability of the transition $i$ to $j$ under action $a$
and $Q_1(b,i)$ is the probability of output $b$ from state $i$.

Let ${M}_2=(S_2,\Sigma,\Delta_2,\{\tilde{P}_a\},Q_2,p_0,s_g)$ be a Moore machine where
$S_2 = S_1 \cup \{\hat{s}\}$, where $\hat{s} \not\in S_1$, and
$\Delta_2 = \Delta_1 \cup \{\hat{b}\}$, where $\hat{b} \not\in \Delta_1$.
For each $a \in \Sigma$, we let 
\begin{equation}
\tilde{P}_a(j,i) = 
	\left\{\begin{array}{ll}
	P_a(j,i) & \mbox{ if $i,j \in S \setminus \{s_g\}$ } \\
	1 & \mbox{ if $i \in \{s_g,\hat{s}\}$ and $j=\hat{s}$ } \\
	0 & \mbox{ otherwise. }
	\end{array}\right.
\end{equation}
Let ${Q}_2(b,i) = Q_1(b,i)$ if $i \in S$ and $b \in \Delta_1$,
$1$ if $i=\hat{s}$ and $b=\hat{s}$, and $0$ otherwise.
We may set $R$ and $\gamma$ arbitrarily.

We observe that $M_1$ reaches its goal state $s_g$ with probability at least $\tau$
if and only if
$M_2$ outputs a string without $\hat{b}$ with probability at most $1-\tau$.
\end{proof}
\end{proposition}

Eisert \etal \cite{emg12} showed that the classical {\sc Non-Occurrence} problem with $\tau=0$ is decidable.

\begin{myproblem}
{{Quantum Non-Occurrence}}
{A quantum Moore machine $M = (S,\Sigma,\Delta,\Phi,\Omega,\rho_0,\Pi)$, and $\tau \in [0,1)$.}
{Are there are $\alpha \in \Sigma^\star$ and $\beta \in \Delta^\star$ so that $\Acc_M(\alpha,\beta) \le \tau$?}
\end{myproblem}

\begin{corollary} 
{\sc Quantum Non-Occurrence} is undecidable.

\begin{proof}
This follows from the undecidability of the classical problem (Proposition \ref{prop:nobs-undecidable}).
\end{proof}
\end{corollary}

For the special case when $\tau = 0$, it is known the classical problem is decidable.
But, in the quantum setting, we have the following result.

\begin{theorem} (Eisert, M\"{u}ller, and Gogolin \cite{emg12}) \\
{\sc Quantum Non-Occurrence} with trivial conditional channel and $\tau=0$ is undecidable.
\end{theorem}


\section{Policy Existence}

In a Markov decision process with finite action space, Blackwell \cite{b65} proved that 
there always exists an optimal policy that is deterministic and stationary.
We describe Blackwell's theorem along with relevant background.

\begin{definition}
Let $\Euc{X}$ be a topological space. 
A collection of subsets of $\Euc{X}$ is called a $\sigma$-algebra of $\Euc{X}$ if it contains
$\Euc{X}$ and it is closed under complementation and countable unions.
The smallest $\sigma$-algebra of $\Euc{X}$ which contains all open subsets of $\Euc{X}$ is called 
the {\em Borel $\sigma$-algebra} of $\Euc{X}$. The elements of this Borel $\sigma$-algebra
are called the {\em Borel subsets} of $\Euc{X}$.
A {\em Borel set} is a Borel subset of a complete separable metric space. 
The set of all bounded real-valued functions over a Borel set $\Euc{S}$ 
is denoted $\Baire(\Euc{S})$.
\end{definition}

Let $M = (\Euc{S},A,P,R,p_0,\gamma)$ be a Markov decision process where 
$\Euc{S}$ is a Borel set, 
$A$ is a finite set of actions, 
$P$ is a conditional probability distribution over $\Euc{S}$ given $\Euc{S} \times A$,
$R \in \Baire(\Euc{S})$ is a reward function,
$p_0 \in \Euc{S}$ is the initial state, 
and $\gamma \in [0,1)$ is a discount factor.

For a positive integer $n$, let $H_n = (\Euc{S} \times A)^{n-1} \times \Euc{S}$ be the history at time $n$.
A policy is a sequence $\pi = (\pi_n)$ where $\pi_n$ is a conditional distribution over $A$ given $H_n$. 
A policy $\pi$ is Markov if each $\pi_n:\Euc{S} \rightarrow A$ is a deterministic map.
A policy $\pi$ is stationary if for all $n$, $\pi_n = f$, for some map $f:\Euc{S} \rightarrow A$.
The value of a policy $\pi$ is given by 
\begin{equation}
\Val_\pi(p_0) = \Exp\left[\sum_{i=0}^{\infty} \gamma^i R(X_i)\right].
\end{equation}
The expectation is taken over a probability space which contains all infinite trajectories
(see Appendix \ref{section:inf-product}).
Here, $(X_n)_{n=0}^{\infty}$, $X_n \in \Euc{S}$, is the sequence of states induced by $\pi$.
Note $\Val_\pi \in \Baire(\Euc{S})$.
A policy $\pi^\star$ is optimal if $\Val_{\pi^\star} \ge \Val_\pi$ for all policy $\pi$.
In this case, we denote the optimal value as $V^\star = \Val_{\pi^\star}$.

\begin{theorem} \label{thm:banach} (Blackwell \cite{b65}, Theorem 5) \\
Let $M = (\Euc{S},A,P,R,p_0,\gamma)$ be a Markov decision process (as defined above).
Suppose the operator $\mathfrak{T}: \Baire(\Euc{S}) \rightarrow \Baire(\Euc{S})$ is defined as
$\mathfrak{T}(V) = \sup_{a \in A} T_a(V)$ where 
\begin{equation}
T_a(V)(p) = R(p) + \gamma \Exp[V(P(p,a))].
\end{equation}
Then, $\mathfrak{T}$ is $\gamma$-Lipschitz, 
that is, $\norm{\mathfrak{T}(U) - \mathfrak{T}(V)} \le \gamma\norm{U-V}$,
so that (by the Banach fixed-point theorem)
$\mathfrak{T}$ has a unique fixed point $V^\star$ and
$\norm{\mathfrak{T}^n(V_0) - V^\star} \le \gamma^n\norm{V_0 - V^\star}$, for all $n$
and any $V_0 \in \Baire(\Euc{S})$.
\end{theorem}

\begin{theorem} \label{thm:blackwell} (Blackwell \cite{b65}) \\
Let $M = (\Euc{S},A,\{P_a\}_{a \in A},R,p_0,\gamma)$ be a Markov decision process (as defined above).
Then:
\begin{enumerate}[(i)]
\item $\pi$ is optimal if and only if $\Val_{\pi} = \sup_{a} T_a(\Val_{\pi})$ 
	(see Theorem 6(f)). 
\item There is an optimal stationary policy for $M$ 
	(see Theorem 7(b)). 
\end{enumerate}
\end{theorem}

Theorems \ref{thm:banach} and \ref{thm:blackwell} show that 
$\pi^\star$ is optimal if its value function $V^\star = \Val_{\pi^\star}$ is a fixed point of $\mathfrak{T}$, 
that is, $V^\star = \mathfrak{T}(V^\star)$.

In what follows, we apply Blackwell's theorems to a quantum observable Markov decision process. 
Given that a quantum observable Markov decision process is a Markov decision process over the set of
density matrices, it suffices to show that the latter forms a Borel space.

\begin{proposition} \label{prop:borel-dens}
Let $\Euc{S}$ be a complex Euclidean space over a finite set $S$.
Then, the set $\Dens(\Euc{S})$ of all density matrices over $\Euc{S}$ is a Borel set.

\begin{proof}
Let $d = |S|$. Then, each element of $\Dens(\Euc{S})$ is a unit trace, positive semidefinite matrix from $M_d(\CC)$. 
The latter is a complete separable metric space (with a norm induced by the inner product $\ip{A}{B} = \Tr(A^\dagger B)$).
To show $\Dens(\Euc{S})$ is Borel, it suffices to show it is closed.
Let $(\rho_n)_{n=0}^{\infty}$ be a convergent sequence in $\Dens(\Euc{S})$ where $\rho_n \rightarrow \eta$ as $n \rightarrow \infty$.
We need to show $\eta \in \Dens(\Euc{S})$.
Since trace is linear, it is continuous; thus, $\Tr(\lim \rho_n) = \lim \Tr(\rho_n) = 1$.
To show $\eta \succeq 0$, consider the map $f(A) = \Tr(Ax_0 x_0^T)$ for some fixed but arbitrary 
$x_0 \in \CC^d$. Since $f$ is linear, it is continuous. Thus, $f(\lim \rho_n) = \lim f(\rho_n)$.
This shows that $f(\eta) = x_0^T \eta x_0 \ge 0$. Since $x_0$ is arbitrary, this shows
$\eta \succeq 0$. Thus, $\eta \in \Dens(\Euc{S})$.
\end{proof}
\end{proposition}

\begin{theorem}
Let $M = (\Euc{S},\Sigma,\Delta,\Lambda,R,\rho_0,\gamma)$ be a quantum observable Markov decision process.
Then, there is an optimal stationary policy for $M$. 

\begin{proof}
Note this follows from Theorem \ref{thm:blackwell} since $\Euc{S}$ is a Borel set (by Proposition \ref{prop:borel-dens}),
$\Sigma$ is finite, $\Lambda$ induces a conditional probability distribution over $\Euc{S}$ given $\Euc{S} \times \Sigma$,
and $R$ is a bounded real-valued function.
\end{proof}
\end{theorem}

Under certain assumptions, the optimal value function has a compact representation
as a piecewise linear and convex function over the probability simplex.
This was originally observed by Sondik \cite{s78}.
In the next theorem, we generalize this observation to the quantum setting.

\begin{theorem} \label{thm:sondik}
Let $M=(\Euc{S},\Sigma,\Delta,\Phi,\Omega,R,\rho_0,\gamma)$ be a quantum observable Markov decision process.
Suppose $\mathfrak{T}: \Baire(\Euc{S}) \rightarrow \Baire(\Euc{S})$ is an operator given by
\begin{equation} \label{eqn:fpo}
\mathfrak{T}(V)(\rho) = \sup_a \left\{ R(\rho) + \gamma\Exp[V(\Phi_a(\rho))] \right\}.
\end{equation}
For any $V_0 \in \Baire(\Euc{S})$ and for any nonnegative integer $n$, 
$V_n = \mathfrak{T}^n(V_0)$ is a piecewise linear and convex function. 

\begin{proof}
Assume that the Kraus decomposition $\Phi_a(\rho) = \sum_i K_{ai}\rho K_{ai}^\dagger$, 
for each quantum channel $\Phi_a$, $a \in \Sigma$.
Also, assume that $\Omega_b(\rho) = L_b\rho L_b^\dagger$, for each operator of the quantum instrument $\Omega$.
Suppose the reward function is given by ${R}(\rho) = \Tr(R\rho) = \ip{R}{\rho}$ for some observable $R$.
For brevity, we have identified the reward operator $R$ with its observable.
Note ${R}$ is bounded since the underlying set of states is finite. 
Then, we have
\begin{eqnarray}
\mathfrak{T}(V)(\rho) 
	& = & \sup_a \left\{ \ip{R}{\rho} + \gamma \sum_b p_b V\left(\frac{\Omega_b \circ \Phi_a(\rho)}{p_b}\right) \right\}.
\end{eqnarray}
Assume inductively that $V$ is piecewise linear and convex. This yields
\begin{eqnarray}
\mathfrak{T}(V)(\rho) 
	& = & \sup_a \left\{ \ip{R}{\rho} + \gamma \sum_b p_b \sup_c \ip{R_c}{\frac{\Omega_b \circ \Phi_a(\rho)}{p_b}} \right\},
\end{eqnarray}
for some set of operators $\{R_c\}$.
Now, using the Kraus forms of $\Phi_a$ and $\Omega_b$, we get
\begin{eqnarray}
\mathfrak{T}(V)(\rho) 
	& = & \sup_a \left\{ \ip{R}{\rho} + \gamma \sum_b \sup_c \ip{\tilde{R}_{abc}}{\rho} \right\},
\end{eqnarray}
where $\tilde{R}_{abc} = \sum_i K_{ai}^\dagger L_b^\dagger R_c L_b K_{ai}$.
This shows that $\mathfrak{T}(V)(\rho) = \sup_d \ \ip{\tilde{R}_d}{\rho}$, for some collection of operators $\{\tilde{R}_d\}$.
Thus, if $\mathfrak{T}$ is applied a finite number of times, the resulting $V$ is
piecewise linear (due to inner product) and convex (due to supremum).
\end{proof}
\end{theorem}

\bigskip

\begin{myproblem}
{{Quantum Policy-Existence}}
{A quantum observable Markov decision process $M$ and $\tau \in \RR$.}
{Is there a policy $\pi$ for $M$ so that $\Val_\pi \ge \tau$?}
\end{myproblem}

\begin{theorem} \label{thm:qopt-undecidable}
{\sc Quantum Policy-Existence} is undecidable.

\begin{proof}
This holds since the corresponding classical problem is undecidable (see Madani \etal \cite{mhc03}, Theorem 4.4).
\end{proof}
\end{theorem}

\begin{myproblem}
{{Quantum Approximate Optimal Policy}}
{A quantum observable Markov decision process $M$ and $\epsilon \in (0,1)$.}
{Is there a policy $\pi$ for $M$ so that $\Val_\pi \ge V^\star - \epsilon$, where $V^\star$ is the optimal value?}
\end{myproblem}

\begin{theorem} \label{thm:qopt-decidable}
{\sc Quantum Approximate Optimal Policy} is decidable.

\begin{proof}
(Sketch)
We apply the standard analytic argument using Theorem \ref{thm:banach}.
Choose an initial $V_0$ and repeat $V_{n+1} = \mathfrak{T}(V_n)$, for $n=0,1,\ldots$,
until $\norm{V_{n+1} - V_n} \le (1-\gamma)\epsilon/\gamma$.
If this halts, we have 
\begin{eqnarray}
\norm{V^\star - V_{n+1}} 
	& \le & \norm{V^\star - \mathfrak{T}(V_{n+1})} + \norm{\mathfrak{T}(V_{n+1}) - V_{n+1}} \\
	& = & \norm{\mathfrak{T}(V^\star) - \mathfrak{T}(V_{n+1})} + \norm{\mathfrak{T}(V_{n+1}) - \mathfrak{T}(V_n)} \\
	& \le & \gamma \norm{V^\star - V_{n+1}} + \gamma \norm{V_{n+1} - V_n},
\end{eqnarray}
which shows $\norm{V_{n+1} - V^\star} \le \epsilon$.
The iteration halts since $\mathfrak{T}$ is a contraction with modulus $\gamma$:
\begin{equation}
\norm{V_{n+1} - V_n}
	= \norm{\mathfrak{T}(V_n) - \mathfrak{T}(V_{n-1})} \le \gamma\norm{V_n - V_{n-1}},
\end{equation}
which shows $\norm{V_{n+1} - V_n} \le \gamma^n \norm{\mathfrak{T}(V_0) - V_0}$.
So, for $n = O(\ln\tfrac{1}{\epsilon})$ we have $\norm{V_{n+1} - V^\star} \le \epsilon$.
Finally, we can check $\norm{V_{n+1} - V_n}$ since the value function is
a compact piecewise-linear and convex map (by Theorem \ref{thm:sondik}).
\end{proof}
\end{theorem}


\section{Concluding remarks}

We may exploit the decoupled structure of our quantum Moore machine to introduce two classical channels
(see Figure \ref{fig:channel}). 
The classical output channel transmits the output of the quantum instrument to the policy algorithm (or agent).
The classical input channel transmits the action chosen by the policy to the quantum conditional channel.

\tikzstyle{boxc}=[draw, minimum size=2em]
\tikzstyle{boxq}=[draw, minimum size=2em]
\tikzstyle{box2}=[draw, minimum size=2em]
\tikzstyle{box3}=[draw, minimum size=2em]
\tikzstyle{init} = [pin edge={to-,thin,double}]
\tikzstyle{exit} = [pin edge={-to,thin,double}]

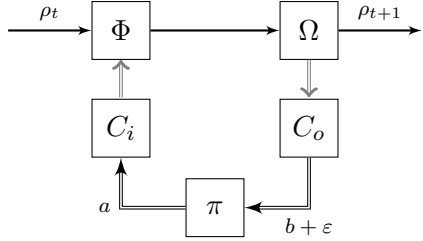
\begin{figure}[h]
\begin{center}
\begin{tikzpicture}[scale=2.0,node distance=2.5cm,auto,>=latex']
    \node [boxc, pin={[init]below:}] (a) {{\color{black} $\Phi$}};
    \node (b) [left of=a,node distance=1.5cm, coordinate] {a};
    \node [boxq, pin={[exit]below:}] (x) [right of=a] {{\color{black} $\Omega$}};
    \node [box2] (ci) [below of=a, node distance=1.3cm] {{\color{black} $C_i$}};
    \node [box2] (co) [below of=x, node distance=1.3cm] {{\color{black} $C_o$}};
    \node [coordinate] (end) [right of=x, node distance=1.5cm]{};
    \path[->,thick] (b) edge node {\scriptsize $\rho_t$} (a);
    \path[->,thick] (a) edge node[name=u] {} (x);
    \draw[->,thick] (x) edge node {\scriptsize $\rho_{t+1}$} (end) ;
    \node [box3, below of=u] (policy) {$\pi$};
    \draw [->, double] (co) |- node [name=bb] {\scriptsize $b+\varepsilon$} (policy);
    \draw [->, double] (policy) -| node [name=aa] {\scriptsize $a$} (ci);
\end{tikzpicture}
\caption{One step in the life-cycle of a quantum Moore machine:
$\Phi$ is a conditional channel, $\Omega$ is a quantum instrument,
$a \in A$ is an input action, $b \in B$ is an output signal,
$\pi$ is the policy, and $C_i,C_o$ are classical (possibly noisy) input and output channels.
}
\label{fig:channel}
\end{center}
\end{figure}

In this work, we assumed that these two channels are noiseless. In a more realistic case, 
one or both of these channels might be noisy.
If the noise characteristics are known, these can be built into the quantum Markov process.
We leave the unknown noise case for future work.


\section*{Acknowledgments}

This material is based on research sponsored 
by Air Force Research Laboratory under agreement number FA8750-18-1-0104. 
The U.S. Government is authorized to reproduce and distribute reprints for Government purposes 
notwithstanding any copyright thereon.
The views and conclusions contained herein are those of the authors and should not be
interpreted as necessarily representing the official policies or endorsements, either
expressed or implied, of Air Force Research Laboratory or the U.S. Government.


\newpage
\appendix

\section{Infinite product spaces} \label{section:inf-product}

Let $S$ be a finite set of states and let $\Euc{S}$ be a complex Euclidean space corresponding to $S$.
Suppose we have a quantum observable Markov decision process $M=(\Euc{S},\Sigma,\Delta,\{\Lambda_a\}_{a \in \Sigma},R,\rho_0,\gamma)$.
Here, we view $M$ equivalently as a Markov decision process $M=(\Dens(\Euc{S}),\Sigma,\{\Lambda_a\}_{a \in \Sigma},R,\rho_0,\gamma)$
whose state space is the set of density matrices $\Dens(\Euc{S})$,
where each $\Lambda_a$ is a conditional probability distribution $P(\Dens(\Euc{S})|\Dens(\Euc{S}) \times \Sigma)$
with support of size at most $|\Delta|$.

Let $\pi = (\pi_n)$ be a policy for $M$.
Given $M$ and $\pi$, consider an infinite sequence $(\rho_n)_{n=0}^{\infty}$ where for each $n$
the random variable  $\rho_n \in \Dens(\Euc{S})$ is distributed according to $\Lambda_{a}(\cdot | \rho_{n-1})$, $a = \pi_n(\rho_{n-1})$.
This infinite sequence of random variables belongs to the probability space of all infinite trajectories.
The latter exists due to the following theorem of Ionescu Tulcea.

\begin{theorem} \label{thm:extension} (Ionescu Tulcea extension theorem (see \cite{s96}, page 249)) \\
Let $(\Omega_n,\mathcal{F}_n)_{n=1}^{\infty}$ be arbitrary measurable spaces
and let $(\Omega,\mathcal{F})$ be their direct product,
where $\Omega = \prod_n \Omega_n$ and $\mathcal{F}$ is the smallest $\sigma$-algebra
containing all the cylinder sets\footnote{These are sets that are trivial in all but a finitely many
positions: $F_{n_1,\ldots,n_k}(B_1,\ldots,B_k) = \{\omega \in \Omega: \omega_{n_j} \in B_j, j=1,\ldots,k\}$,
where $B_1 \in \mathcal{F}_1,\ldots,B_k \in \mathcal{F}_k$.}.
Suppose that a probability measure $P_1$ is given on $(\Omega_1,\mathcal{F}_1)$
and that for every set $(\omega_1,\ldots,\omega_n) \in \Omega_1 \times \ldots \times \Omega_n$,
for $n \ge 1$, probability measures $P(\cdot | \omega_1,\ldots,\omega_n)$
are given on $(\Omega_{n+1},\mathcal{F}_{n+1})$.
Suppose that for every $B \in \mathcal{F}_{n+1}$, the functions $P(B | \omega_1,\ldots,\omega_n)$
are Borel functions on $(\omega_1,\ldots,\omega_n)$ and let
\begin{equation}
P_n(A_1 \times \ldots \times A_n) =
	\int_{A_1} P_1(\omega_1) \int_{A_2} P(d\omega_2 | \omega_1)
	\ldots
	\int_{A_n} P(d\omega_n | \omega_1,\ldots,\omega_{n-1}),
\ \ \
\mbox{ for $A_i \in \mathcal{F}_i$, $n \ge 1$.}
\end{equation}
Then, there is a unique probability measure $P$ on $(\Omega,\mathcal{F})$ so that
\begin{equation}
P(\{\omega : \omega_1 \in A_1,\ldots,\omega_n \in A_n\}) = 
	P_n(A_1 \times \ldots \times A_n),
\ \ \
\mbox{ for every $n \ge 1$}
\end{equation}
and there is a random sequence $X = (X_1(\omega),X_2(\omega),\ldots)$ so that
\begin{equation}
P(\{\omega : X_1(\omega_1) \in A_1,\ldots,X_n(\omega) \in A_n\}) = 
	P_n(A_1 \times \ldots \times A_n),
\ \ \
\mbox{ where $A_i \in \mathcal{F}_i$.}
\end{equation}
\end{theorem}

\medskip
Let $(\Omega,\mathcal{F},P)$ be the unique probability space for the set of all infinite trajectories 
induced by the quantum Markov process (which exists due to Theorem \ref{thm:extension}).
The value function of policy $\pi$ is given by
\begin{equation} \label{eqn:value-function}
\Val_\pi(\rho_0) := \Exp\left[ \sum_{i=0}^{\infty} \gamma^i R(\rho_i) \right]
\end{equation}
where the expectation is taken with respect to the probability measure $P$.

\ignore{
If the expectation and the limiting process commute in \eqref{eqn:value-function}, then
\begin{equation}
\Exp\left[ \lim_{n \rightarrow \infty} \sum_{i=0}^{n} \gamma^i R(\rho_i) \right]
=: \Val_\pi =
\lim_{n \rightarrow \infty} \Exp\left[ \sum_{i=0}^{n} \gamma^i R(\rho_i) \right].
\end{equation}
This holds due to the following theorem.
}

The next theorem of Lebesgue allows us to interchange the order of the expectation
with the limiting process in \eqref{eqn:value-function}.

\begin{theorem} \label{thm:lebesgue}
(Lebesgue dominated convergence theorem (see \cite{s96}, page 187)) \\
Let $\eta$, $Z$, $Z_1,Z_2, \ldots$ be random variables such that 
$|Z_n| < \eta$, 
$\Exp[\eta] < \infty$
and $Z_n \rightarrow Z$ (almost surely).
Then, $\Exp[|Z|] < \infty$, $\Exp[Z_n] \rightarrow \Exp[Z]$ and $\Exp[|Z_n - Z|] \rightarrow 0$, 
as $n \rightarrow \infty$.
\end{theorem}

By Theorem \ref{thm:lebesgue}, we have that
\begin{equation}
\Exp\left[ \lim_{n \rightarrow \infty} \sum_{i=0}^{n} \gamma^i R(\rho_i) \right]
=
\lim_{n \rightarrow \infty} \Exp\left[ \sum_{i=0}^{n} \gamma^i R(\rho_i) \right].
\end{equation}

\section{Example: Bloch transducer}

For a Pauli matrix $A$, let $R_A(\theta) = \exp(-\ii A/2)$ be the rotation matrix about the $A$ axis
(see \cite{nc00}, page 174). We have
\begin{equation}
R_{x}(\theta) 
	= \begin{pmatrix} 
			\cos\tfrac{\theta}{2} & -\ii\sin\tfrac{\theta}{2} \\
			-\ii\sin\tfrac{\theta}{2} & \cos\tfrac{\theta}{2} 
			\end{pmatrix},
R_{y}(\theta) 
	= \begin{pmatrix} 
			\cos\tfrac{\theta}{2} & -\sin\tfrac{\theta}{2} \\
			\sin\tfrac{\theta}{2} & \cos\tfrac{\theta}{2}
			\end{pmatrix},
R_{z}(\theta) 
	= \begin{pmatrix} 
			e^{-\ii\theta/2} & 0 \\
			0 & e^{\ii\theta/2}
			\end{pmatrix}
\end{equation}

Let $M=(\Dens(\Euc{S}),\{a_1,a_2\},\{b_{-1},b_1\},\Phi,\{\Omega_{b_{\pm 1}}\},\rho_0,\Pi)$
be a quantum Moore machine, where
$\Euc{S} = \vspan\{\ket{0},\ket{1}\}$, $\rho_0 = \ketbra{+}{+}$ and
$\Pi$ is the projection onto the diagonal density matrices (probability simplex).
The conditional channel $\Phi$ is given by the following two quantum channels:
\begin{equation}
\Phi_{a_1}(\rho) = \tfrac{1}{2}\rho + \tfrac{1}{2}R_x(\tfrac{\pi}{3})\rho R_x(\tfrac{\pi}{3})^\dagger,
\ \ \
\Phi_{a_2}(\rho) = \tfrac{1}{2}\rho + \tfrac{1}{2}R_y(\tfrac{\pi}{3})\rho R_y(\tfrac{\pi}{3})^\dagger.
\end{equation}
The quantum instrument $\Omega$ is given by the two Kraus operators:
\[
    \Omega_{b_{\pm 1}}(\rho) = \tfrac{1}{2}R_z(\tfrac{\pm\pi}{3})\rho R_z(\tfrac{\pm\pi}{3})^\dagger.
\]

\begin{figure}[h]
\begin{center}
\includegraphics[scale=0.25]{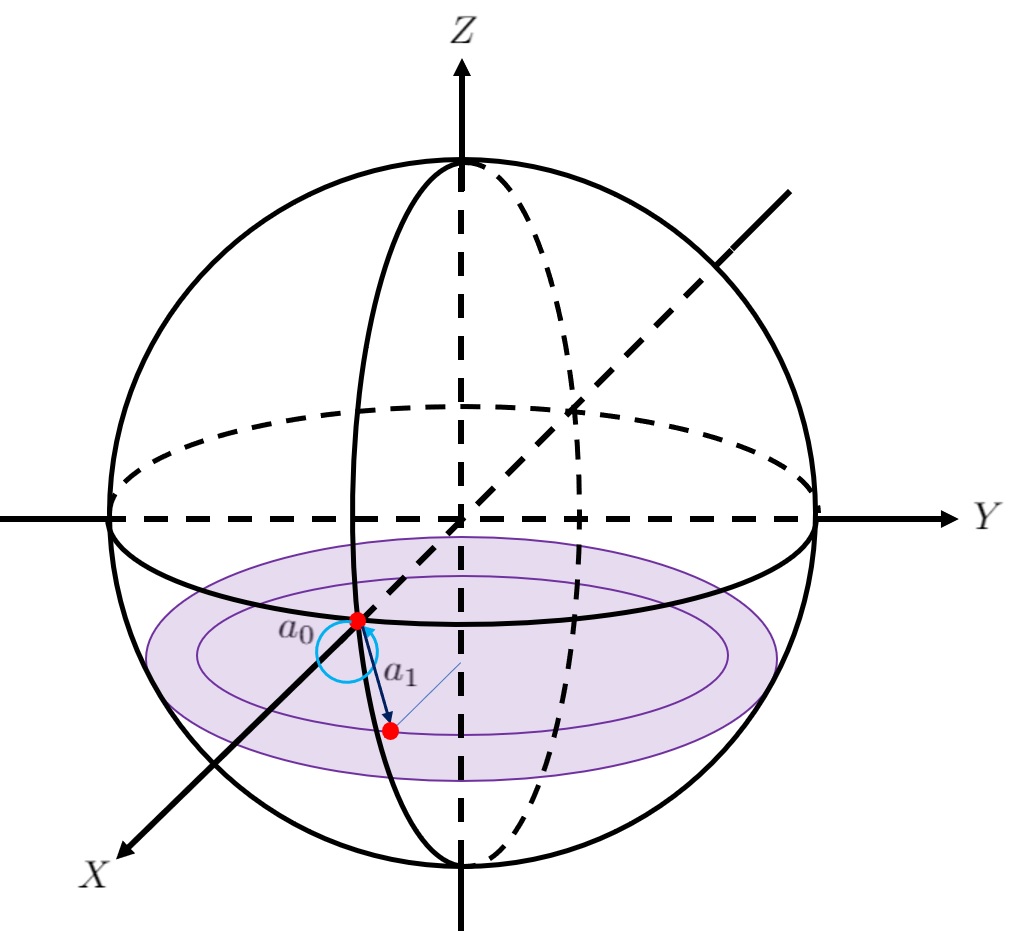}
\includegraphics[scale=0.25]{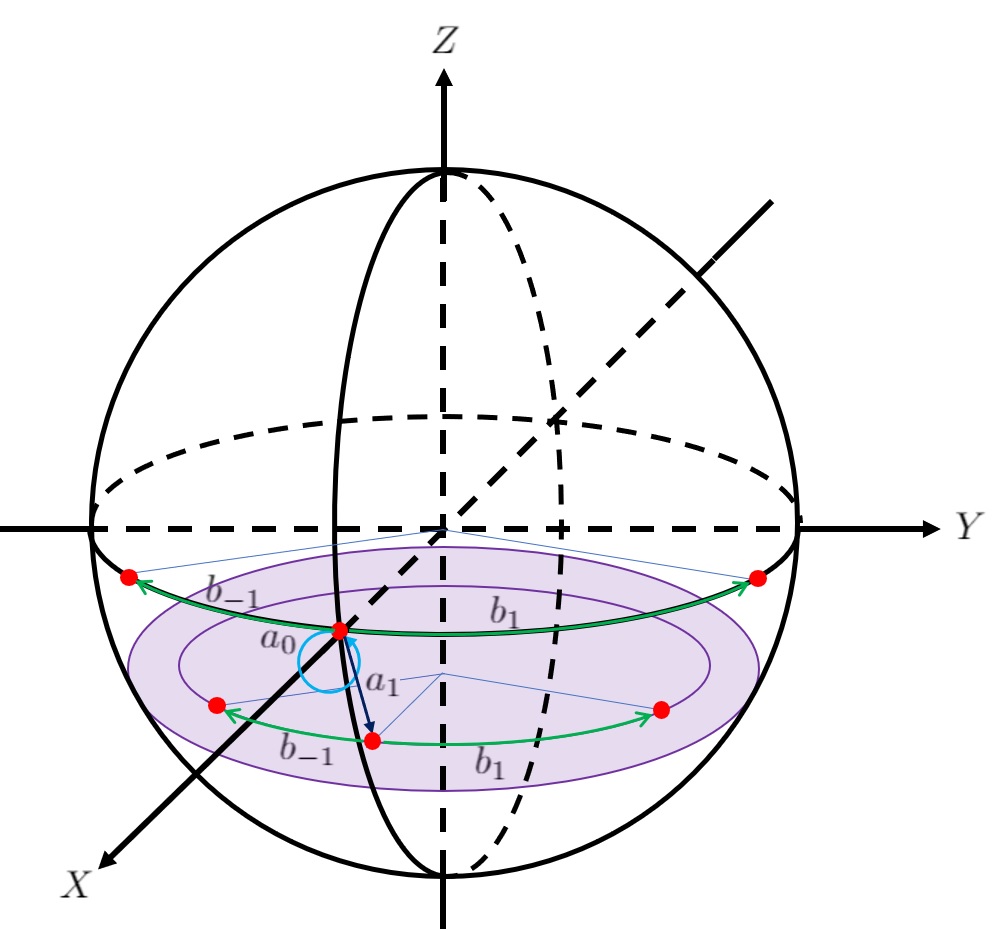}
\includegraphics[scale=0.25]{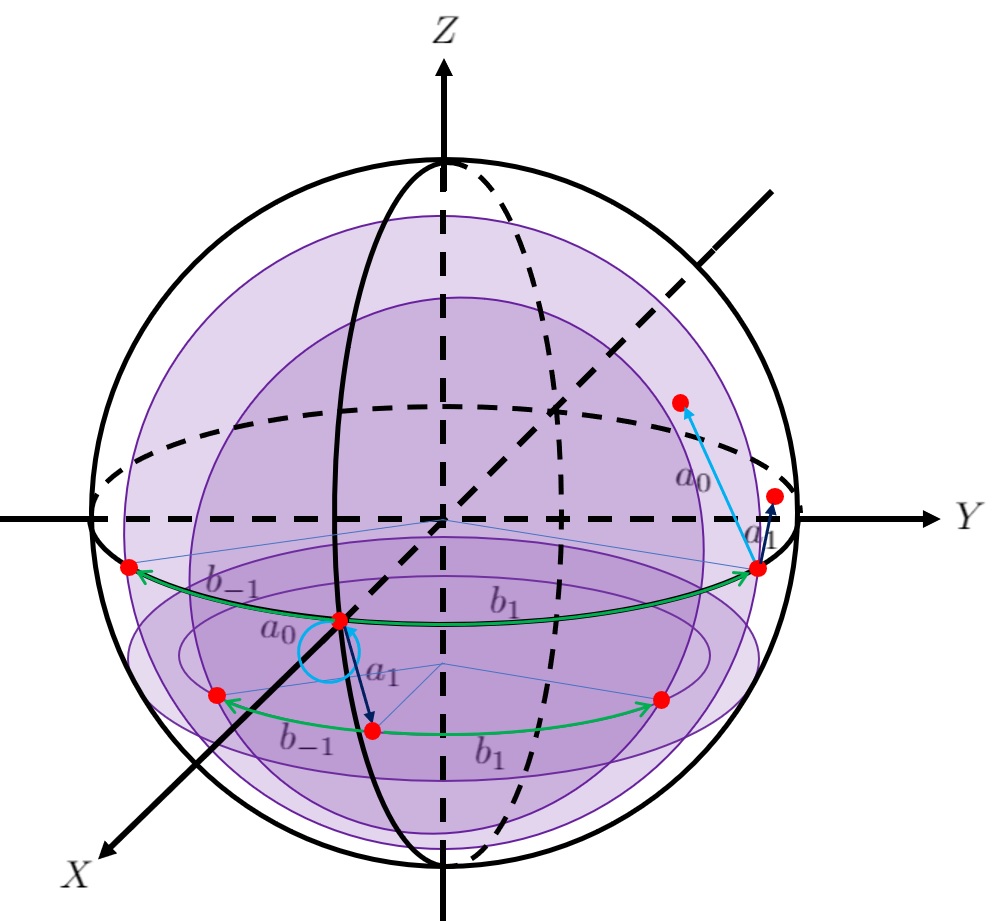}
\caption{Example: three steps in the life-cycle of a quantum transducer.}
\end{center}
\end{figure}



\begin{thebibliography}{1}

\bibitem{bba14}
{\sc J. Barry, D. Barry, S. Aaronson}.
\newblock ``Quantum partially observable Markov decision processes.''
\newblock {\em Physical Review A} {\bf 90}:032311, 2014.

\bibitem{bs78}
{\sc D. Bertsekas, S. Shreve}.
\newblock {\em Stochastic Optimal Control: The Discrete-Time Case}.
\newblock Academic Press, 1978.

\bibitem{b65}
{\sc D. Blackwell}.
\newblock ``Discounted Dynamic Programming.''
\newblock {\em Annals of Mathematical Statistics} {\bf 36}:226-235, 1965.

\bibitem{bjkp05}
{\sc V. Blondel, E. Jeandel, P. Koiran, N. Portier}.
\newblock ``Decidable and Undecidable Problems about Quantum Automata.''
\newblock {\em SIAM J. Computing}, {\bf 34}(6):1464-1473, 2005.

\bibitem{emg12}
{\sc J. Eisert, M. M\"{u}ller, C. Gogolin}.
\newblock ``Quantum measurement occurrence is undecidable.''
\newblock {\em Physical Review Letters} {\bf 108}:260501, 2012.

\bibitem{hu79}
{\sc J. Hopcroft, J. Ullman}.
\newblock {\em Introduction to Automata Theory, Languages, and Computation}.
\newblock Addison-Wesley, 1979.

\bibitem{mhc03}
{\sc O. Madani, S. Hanks, A. Condon}.
\newblock ``On the undecidability of probabilistic planning and related stochastic optimization problems.''
\newblock {\em Artificial Intelligence} {\bf 147}:5-34, 2003.

\bibitem{mbw11}
{\sc A. Monras, A. Beige, K. Wiesner}.
\newblock ``Hidden Quantum Markov Models and Non-Adaptive Read-Out of Many-Body States.''
\newblock {\em Appl. Math. Comp. Sciences} {\bf 3}:93, 2011.

\bibitem{nc00}
{\sc M. Nielsen, I. Chuang}.
\newblock {\em Quantum Information and Quantum Computation}.
\newblock Cambridge University Press, 2000.

\bibitem{rn10}
{\sc S. Russell, P. Norvig}.
\newblock {\em Artificial Intelligence: A Modern Approach}, 3rd edition.
\newblock Prentice Hall, 2010.

\bibitem{s96}
{\sc A. Shiryaev}.
\newblock {\em Probability}, second edition.
\newblock Springer, 1996.

\bibitem{s78}
{\sc E.J. Sondik}.
\newblock ``The Optimal Control of Partially Observable Markov Processes over the Infinite Horizon: Discounted Costs.''
\newblock {\em Operations Research} {\bf 26}(2): 282-304, 1978.

\bibitem{s04}
M. Szegedy.
\newblock ``Quantum Speed-Up of Markov Chain Based Algorithms.''
\newblock Proc. 45th Symposium on Foundations of Computer Science, 32-41, 2004.

\bibitem{sb18}
{\sc R. Sutton, A. Barto}.
\newblock {\em Reinforcement Learning: An Introduction}, second edition.
\newblock {The MIT Press}, 2018.

\bibitem{tkrwv10}
{\sc K. Temme, M. Kastoryano, M.B. Ruskai, M. Wolf, F. Verstraete}.
\newblock ``The $\chi^2$-divergence and mixing times of quantum Markov processes.''
\newblock {\em Journal of the Mathematical Physics} {\bf 51}, 122201, 2010.

\bibitem{wc08}
{\sc K. Wiesner, J. Crutchfield}.
\newblock ``Computation in finitary stochastic and quantum processes.''
\newblock {\em Physica D}, {\bf 237}:1173-1195, 2008.

\bibitem{w18}
{\sc J. Watrous}.
\newblock {\em The Theory of Quantum Information}.
\newblock Cambridge University Press, 2018.

\bibitem{w17}
{\sc M. Wilde}.
\newblock {\em Quantum Information Theory}, second edition.
\newblock Cambridge University Press, 2017.

\bibitem{wcp11}
{\sc M. Wolf, T. Cubitt, D. Perez-Garcia}.
\newblock ''Are problems in quantum information theory (un)decidable?''
\newblock arXiv:1111.5425 [quant-ph].


\end{thebibliography}
\end{document}